\documentclass{an}
\usepackage{graphicx}
\usepackage{psfig}
\usepackage{times}
\begin{document}
\lhead[\thepage]{R. Neuh\"auser et al.: 
Imaging and spectroscopy of DENIS-P~J104814$-$395606}
\rhead[Astron. Nachr./AN~{\bf XXX} (200X) X]{\thepage}
\headnote{Astron. Nachr./AN {\bf 32X} (200X) X, XXX--XXX}

\title{Deep infrared imaging and spectroscopy of the nearby late M-dwarf 
DENIS-P~J104814$-$395606\thanks{Data presented here were obtained in ESO
programs 66.D-0135(A), 66.C-0138(B), 67.C-0325(A), 67.C-0514(A), and 67.C-0354(A)
on La Silla and Cerro Paranal, as well as with the 2.2m telescope of the 
University of Hawai'i Institute for Astronomy on Mauna Kea.}}

\author{R. Neuh\"auser\inst{1}, E.W. Guenther\inst{2}, J. Alves\inst{3}, 
N. Grosso\inst{1}, Ch. Leinert\inst{4}, Th. Ratzka\inst{4}, Th. Ott\inst{1}, 
M. Mugrauer\inst{1}, F. Comer\'on\inst{3}, A. Eckart\inst{5} \and W. Brandner\inst{3}}
\institute{
MPI f\"ur extraterrestrische Physik, D-85740 Garching, Germany
\and 
Th\"uringer Landessternwarte Tautenburg, Sternwarte 5, D-07778 Tautenburg, Germany
\and 
European Southern Observatory, Karl-Schwarzschild-Stra\ss e 2, D-85748 Garching, Germany
\and
MPI f\"ur Astronomie, K\"onigstuhl, D-69117 Heidelberg, Germany
\and
I. Physikalisches Institut, Universit\"at zu K\"oln, Germany}

\date{Received {\it date will be inserted by the editor}; 
accepted {\it date will be inserted by the editor}} 

\abstract{We obtained deep H- and K-band images of DENIS-P~J104814$-$395606
using SofI and the speckle camera SHARP-I at the ESO-3.5m-NTT as well as QUIRC
at the Mauna Kea 2.2m telescope between December 2000 and June 2001.
The target was recently discovered as nearby M9-dwarf among DENIS sources (Delfosse et al. 2001).
We detect parallactic motion on our images and determine the distance to be $4.6 \pm 0.3$ pc,
more precise than previously known.
From the available colors, the distance, and the spectral type,
we conclude from theoretical models that the star has
a mass of $\sim 0.075$ to $0.09~$M$_{\odot}$ and an age of $\sim 1$ to 2 Gyrs.
We also obtained H- and K-band spectra of this star with ISAAC at the VLT.
A faint companion candidate is detected 6$^{\prime \prime}$ NNW of the star, 
which is $6.4 \pm 0.5$ mag fainter in H. 
%
%
However, according to another image taken several month later,
the companion candidate is not co-moving with the M9-dwarf. 
Instead, it is a non-moving background object.
Limits for undetected companion candidates are such that we can exclude
any stellar companions outside of $\sim 0.25^{\prime \prime}$ (1 AU),
any brown dwarf companions (above the deuterium burning mass limit)
outside of $\sim 2^{\prime \prime}$ (9 AU), and also any companion
down to $\sim 40$~M$_{\rm jup}$ with $\ge 0.15^{\prime \prime}$ (0.7 AU)
separation, all calculated for an age of 2 Gyrs. Our observations show that 
direct imaging of sub-stellar companions near the deuterium burning mass limit 
in orbit around nearby ultra-cool dwarfs is possible, 
even with separations that are smaller than the semi-major axis of the
outermost planet in our solar system, namely a few tens of AU. 
%
%
\keywords{Editorial notes -- instruction for authors}
}

\correspondence{rne@mpe.mpg.de}

\maketitle

\section{Introduction}

The object DENIS-P~J104814$-$395606 (hereafter Denis~1048$-$39)
was discovered as very red source in the DEep Near-Infrared Survey (DENIS,
Epchtein 1997) by Delfosse et al. (2001):
I=$12.67 \pm 0.05$, J=$9.59 \pm 0.05$, and K$_{\rm s}=8.58 \pm 0.05$ mag.
Together with its USNO B- and R-band colors (B=$19.0 \pm 0.3$ and R=$15.7 \pm 0.2$ mag),
it was classified as late M-type star; a follow-up spectrum taken with 
Keck showed the spectral type to be M9; from the previous USNO and DENIS imaging, 
a high proper motion of $1.516 \pm 0.012^{\prime \prime}$ per year to the SE was determined;
from its magnitudes and spectral type, a distance of $4.1 \pm 0.6$ pc was determined;
its galactic UVW velocities are typical for the thin disk (Delfosse et al. 2001).
Deacon \& Hambly (2001) obtained its trigonometric parallax from archived
plates to be $192 \pm 37$ mas, i.e. a distance of $5.2 ^{+ 1.2} _{-0.8}$ pc,
still consistent with the photometric distance estimate from Delfosse et al.
(2001) for a single or binary star.

Because of its small distance and intrinsically faint 
magnitude, such an object is
a promising target for direct imaging detection of sub-stellar companions,
both brown dwarfs and giant planets. Due to the problem of dynamic range, namely
that sub-stellar objects are too faint and too close to their primary stars,
it is difficult to directly detect such objects in orbit around a normal star.
One way around this problem is to observe intrinsically faint objects like
so-called ultra-cool dwarfs (i.e. dwarfs with spectral type M6 or later, 
including brown dwarfs). If they are also very nearby, one can probe
separations which are of the order of the semi-major axes of the giant planets of
our own solar system.

So far, only a few brown dwarf companion candidates detected directly in orbit
around normal stars are confirmed by both spectroscopy and proper motion, 
the first of which was Gl~229~B (Nakajima et al. 1995, 
Oppenheimer et al. 1995), and the youngest of which is TWA-5~B
(Lowrance et al. 1999, Neuh\"auser et al. 2000).
The first four confirmed brown dwarf companions\footnote{in addition to 
Gl~229~B and TWA-5~B mentioned above, there are G~196-3~B (Rebolo et al. 1998)
and GD~570~D (Burgasser et al. 2000)} all orbit M-type dwarf stars.
Only the fifth such companion, HR~7329~B, orbits an A0-type star
(Lowrance et al. 2000, Guenther et al. 2001).
This may indicate that nature prefers binaries with not too different masses,
another motivation to search for sub-stellar companions around M-type stars;
however, it could also be a selection effect, because companions with very
different masses and magnitudes are much more difficult to detect.

In Sect. 2, we show our first two deep infrared (IR) imaging observation of Denis~1048$-$39.
Follow-up H- and K-band spectroscopy is presented in Sect. 3.
Two more follow-up images of Denis~1048$-$39 are used 
to derive the proper motion of the companion candidate in Sect. 4
and the parallax of Denis~1048$-$39 itself in Sect. 5. 
Then, in Sect. 6, we search for closer companions with the speckle 
camera SHARP-I. We discuss and summarize our results in Sect. 7.

\section{The first two deep imaging observations}

Deep imaging with high dynamic range can best be done in the near IR,
because of low magnitude difference between the primary and possible sub-stellar
companions, good sky conditions like seeing (compared to, e.g., the optical), and 
high sensitivity of near IR detectors (compared to, e.g., the thermal IR).

We observed Denis~1048$-$39 for the first time with the Son of Isaac 
(SofI\footnote{see www.ls.eso.org/lasilla/Telescopes/NEWNTT/}) at the 3.5m
New Technology Telescope (NTT) of the European Southern Observatory (ESO) on La Silla,
Chile, during the morning twilight of 8 December 2000 (at airmass 1.0), while the targets 
of that program (companions to nearby neutron stars) were not observable anymore.
We used the small SofI field with 0.147$^{\prime \prime}$ per pixel
for better angular resolution and obtained $30 \times 30$ sec integrations.
Darks, flats, and standards were observed in the same night with the same set-up
and data reduction was 
done with {\em eclipse}\footnote{see www.eso.org/projects/aot/eclipse/}.
We measured a FWHM of 0.51$^{\prime \prime}$ in our Denis~1048$-$39 images, 
see Table 1 for the observations log.

%
%
Using the faint HST standard stars S427-D and S341-D, we obtain 
the as yet unknown H-band magnitude for Denis~1048$-$39, namely
H=$9.55 \pm 0.16$ mag. This value is the mean of three measurements
taken with SofI (see Table 1 for the observation log), which individualy show 
relatively large errors ($\sim 0.15$ mag each) and a large scatter (9.67, 9.37, 
and 9.60 mag, still consistent with being constant), 
because we observed only few photometric standard stars 
(aiming at pure detections of companion candidates rather
than precise photometry), so that the mean of the three measurements
also has a relatively large error, even though the star is quite bright.
Our H-band magnitude
is roughly consistent with its previously known BRIJK colors and 
its spectral type M9, as well as a distance of roughly 4 to 5 pc 
(M$_{\rm H} \simeq 10.9$ mag for few Gyr old M9 dwarf stars
according to Kirkpatrick \& McCarthy 1994).

\begin{figure}
\vbox{\psfig{figure=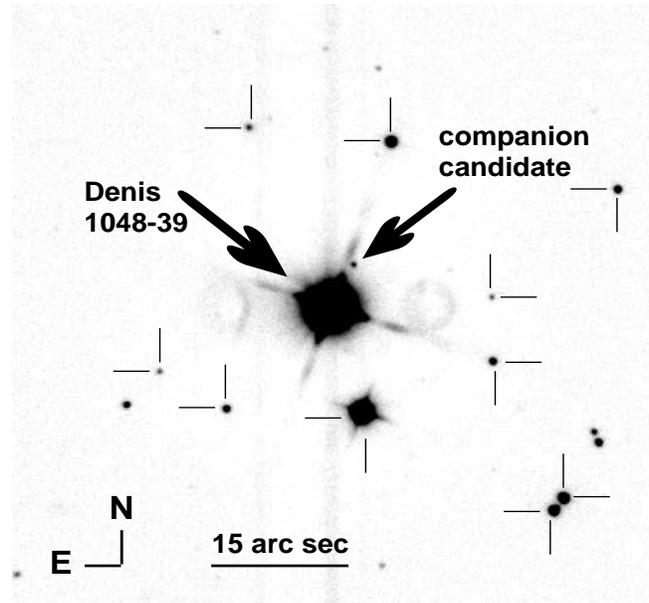,width=8.5cm,height=8cm}}
\caption{NTT/SofI H-band image of Denis~1048$-$39 with surrounding field,
taken on 8 Dec 2000. The faint companion candidate is clearly seen 
$\sim 6 ^{\prime \prime}$ NNW of the primary.
The other marked stars are used as background comparison stars
for the measurement of proper and parallactic motion.}
\end{figure}

We detected a faint object near Denis~1048$-$39 (Fig. 1) with
H=$15.9 \pm 0.5$ mag, i.e. $6.4 \pm 0.5$ mag fainter than Denis~1048$-$39.
In December 2000, the companion candidate was located 
$5.766 \pm 0.014^{\prime \prime}$ NNW of Denis~1048$-$39 (see Table 2).

\begin{table}
\begin{tabular}{llrlc}
\multicolumn{5}{c}{\bf Table 1. Observations log} \\ \hline
Telecope & instrument & obs. date & filter & FWHM \\ \hline
NTT$^{1}$  & SofI  & 8 Dec 2000 & H & 0.51$^{\prime \prime}$ \\
2.2m$^{2}$ & QUIRC & 15 Mar 2001 & H & 1.20$^{\prime \prime}$ \\
NTT$^{1}$  & SofI  & 6 Apr 2001 & H$^{4}$ & 1.36$^{\prime \prime}$ \\
VLT$^{3}$  & ISAAC & 21 Apr 2001 & H \& K$^{5}$ & 0.40$^{\prime \prime}$ \\
NTT$^{1}$  & SofI  & 9 Jun 2001 & H & 0.70$^{\prime \prime}$ \\ 
NTT$^{1}$  & SHARP-I  & 2 Jul 2001 & K & 0.34$^{\prime \prime}$ \\ \hline
\end{tabular}
Remarks: (1) La Silla. (2) Mauna Kea. (3) Cerro Paranal.
(4) Companion candidate not detected due to insufficient dynamic range.
(5) Companion candidate not detected in aquisition image because of
very short integration time (1.77 sec).
\end{table}

%
%

If the companion candidate would be bound, it should be a sub-stellar object.
Whether the candidate indeed is bound, can be check best with a 2nd epoch
image to measure its proper motion.

We obtained our 2nd epoch image with the 
QUick InfraRed Camera (QUIRC\footnote{see http://www.ifa.hawaii.edu/88inch/}),
a $1k \times 1k$ Hawaii array, at the 2.2m (88 inch) telescope of the 
University of Hawai'i Institute of Astronomy on Mauna Kea 
in the nights of March 13/14 and 14/15. 
The images of both nights were flat fielded, sky subtracted, shifted, 
and co-added for the two nights separately 
with {\em MIDAS}\footnote{see www.eso.org/projects/esomidas/}.
Because the seeing was better in the second of those two nights, we 
use only the data of that night
%
%
(174 co-added images with 5 sec integration time each).
Because of the relatively large airmass ($\sim 2$) of Denis~1048$-$39 
for our Mauna Kea observation, the FWHM is not very good anyway (Table 1).

The companion candidate is marginaly detected in the QUIRC image.
Pixel scale (0.216$^{\prime \prime}$ per pixel) and orientation 
(north is up, but tilted by $5.5 ^{\circ}$ to the left)
of the detector QUIRC were measured with our image of the binary star GJ 490 
obtained in the same night with QUIRC (compared to its 2MASS 
image\footnote{see www.ipac.caltech.edu/2mass/}).
The separation between primary object and companion candidate 
($6.0 \pm 0.3 ^{\prime \prime}$) in this 2nd epoch image 
is not inconsistent with the 1st epoch image (Fig. 1), 
so that it could be a co-moving pair,
even though the significance is low because of the large FWHM
in the 2nd epoch image.

\section{Spectroscopic follow-up observation}

%
%

H- and K-band spectra were taken with the Infrared
Spectrograph and Array Camera (ISAAC) at the ESO 8.2m telescope Antu,
Unit Telescope No. 1 (UT1) of the Very Large Telescope (VLT) on Cerro Paranal.
Both spectra were obtained on 21 April 2001 in service mode and consist
of 40 co-added 60 sec exposures each through a $1^{\prime \prime}$ slit.
The slit was aligned neither along the position angle of the primary and its companion 
candidate, nor perpendicular to that angle, but in between those two angles,
so that we could detect the signal from the companion candidate in the wing of
the PSF of the primary: If the slit would have been aligned along the position angle, 
the primary would have been too bright, and if we would have aligned the slit 
perpendicular to that angle, the signal of the faint companion candidate
would have been on top of the bright signal from the primary in the collapsed spectral 
profile. An angle in between the two extremes is the best compromise according to our 
experience (e.g. Neuh\"auser et al. 2001). Darks, flats, and arcs were taken in the same night.

The target aquisition was done in the following way:
First, a short aquisition image was taken; then, the slit was aligned 
in north-south direction (as explained above) and centered on the primary;
then, a small offset to the west was applied as given by the separation
of the companion candidate observed in Dec 2000.
Such an aquisition sequence with the VLT is usually very precise and accurate.
If the companion candidate would indeed be bound, it should have the
same proper and parallactic motion as the primary.

After standard data reduction (dark subtraction, normalization, flat fielding, 
sky subtraction, wavelength calibration, and co-adding the spectra), 
we have searched for the faint signal from the companion candidate in the
wing of the bright primary's PSF. The separation between the peak of the 
primary in the slit and the peak of the companion candidate should be 35 pixel
(i.e. 5.2$^{\prime \prime}$), the separation projected onto the slit orientation.
However, we could not find the faint companion candidate's signature.
This can be taken as first evidence for the companion candidate to be
an unrelated, namely non-moving, background object:
If it would have been a co-moving companion, it should have been in the 
slit\footnote{The same argument, the other way around, was used by Lowrance et al. (2000)
for the HR~7329~B companion candidate: Because it was in the slit after blind offset,
it is probably a co-moving companion, which was later confirmed by Guenther et al. (2001).}.

The H- and K-band spectra, which we obtained for the Denis~1048$-$39 primary,
even though located slightly outside the slit, are typical for an M9 dwarf
as far as the general shape of the continuum and the presence and
strength of absorption lines (e.g. Na and CO) are concerned.

%
%
%

\section{Proper motion of the companion candidate}

Because the image obtained at Mauna Kea has a large FWHM, we had to take
another image to measure the proper motion with higher significance.
We tried to obtain such an image of Denis~1048$-$39 and its companion candidate
on 6 April 2001 with SofI at the NTT (at airmass 1.2), while the targets of that program
(in the $\rho$ Ophiuchi dark cloud) were not visible at good airmass. 
However, due to unfavorable seeing of $\sim 1.4^{\prime \prime}$, 
the companion candidate is not detected in the those images 
($30 \times 30$ sec each in H with the small SofI field, Table 1).
The Denis~1048$-$39 primary itself and several other background comparison
stars are well detected in this image, which we will use below
for the determination of the parallax of Denis~1048$-$39.

Then, we took another image on 9 June 2001 (at airmass 1.4), again with Sofi 
at the NTT, while the targets of that program (dark clouds) were too low
(again $30 \times 30$ sec in H with the small SofI field, Table 1).
This image is useful to measure the proper motion of the companion 
candidate (by comparison with the 1st image, Table 2) and also to determine the 
parallax of the Denis~1048$-$39 primary -- together with the previous two SofI 
images obtained in $\sim 1/4$ year intervalls, see Sect. 5.

\begin{table}
\begin{tabular}{lccc}
\multicolumn{4}{c}{\bf Table 2. Separations of the companion candidate} \\ \hline
Obs. date & $\Delta \alpha$ & $\Delta \delta$ & separation \\ \hline
8 Dec 2000 & $2.510(24)^{\prime \prime}$ W & $5.191(13)^{\prime \prime}$ N & $5.766(14)^{\prime \prime}$ \\ 
9 Jun 2001 & $1.517(25)^{\prime \prime}$ W & $5.432(21)^{\prime \prime}$ N & $5.640(18)^{\prime \prime}$ \\ \hline
\end{tabular}
\end{table}

The two separations measured between the primary and the companion candidate 
as obtained from the 1st and 4th image (both with SofI) are listed in Table 2.
While the errors in $\Delta \alpha$ and $\Delta \delta$ given in Table 2
take into account an uncertainty in the orientation of the detector (to the north)
of $\pm 0.01^{\prime \prime}$, the error in separation does not include that error,
because the separation can be measured independant of the orientation.
All errors given in Table 2 include an uncertainty in the pixel scale of $1~\%$.

The separation between the primary and the companion candidate 
has changed significantly from Dec 2000 and June 2001.
This cannot be explained by orbital motion after half a year. 
Hence, the two objects do not form a common proper motion pair. 
The companion candidate is clearly a background object.

Assuming that the companion candidate is a non-moving background object,
the observed motion of Denis~1048$-$39 itself (to the SW) relative to this 
background object from Dec 2000 to June 2001 (Table 2) is consistent with 
its known proper motion (Delfosse et al. 2001), taking into account some
parallactic motion of Denis~1048$-$39 (see below).

\section{The parallax of Denis~1048$-$39}

For determining the parallax of Denis~1048$-$39, we measured the 
separations in $\alpha$ and $\delta$ between Denis~1048$-$39 and ten
comparison stars (marked in Fig. 1) including all separations between 
comparison stars, in the same way for all three images taken with SofI.
First, we checked
that the separations between the comparison stars did not change.
The precision in the Gaussian centering of the comparison stars and the target
ranges from 0.16 pixel (24 mas) in the best image (Dec 2000) to 0.33 pixel (49 mas) in
the April 2001 image. Hence, the precision in one measurement of a parallactic 
displacement (i.e. the difference between two stars from one image to another one) 
ranges from 0.5 to 0.7 pixel, i.e. 74 to 103 mas.
When we average the relative motion (from Dec 2000 to April 2001)
between the target and several comparison stars, 
the precision of the target motion is again better, 
namely $\sim 33$ mas in any one direction.

The (averaged) motion of Denis~1048$-$39 with respect to the comparison stars from
Dec 2000 to Apr 2001 is $0.750 \pm 0.043 ^{\prime \prime}$ to the west
and $0.321 \pm 0.101 ^{\prime \prime}$ to the south.
From this motion, we have to subtract the proper motion 
(known from Delfosse et al. 2001, and Deacon \& Hambly 2001 to be
$\mu \alpha = -1.147 \pm 0.008^{\prime \prime}$ and 
$\mu \delta = -0.992 \pm 0.009^{\prime \prime}$).
The remaining parallactic motion from Dec to April (0.326 yrs) is 
$0.376 \pm 0.043 ^{\prime \prime}$ to the west
and $0.002 \pm 0.101 ^{\prime \prime}$ to the north.
Similarily, the parallactic motion from April to June (0.175 yrs) is
$0.151 \pm 0.062 ^{\prime \prime}$ to the west 
and $0.141 \pm 0.023 ^{\prime \prime}$ to the north.

%
%

One can estimate the parallactic motion for each position in the sky and
any set of observing dates. For the Denis~1049$-$39 position and our observing dates,
we can estimate the parallactic motion in right ascension and declination for
a range of possible distances, e.g. from 4 to 6 pc. Those estimates have to 
be compared with the observed motion to derive the true distance.

In Fig. 2, we show the {\em difference} between expected and observed parallactic 
motion (in $\alpha$, $\delta$, and total) for a range in distance. 
The total motion (Dec 2000 to June 2001) is best consistent with 
a distance of $\sim 4.6$ pc. Given our astrometric precision of $\sim 33$ mas 
(see above), the error is $\pm 0.3$~pc, as seen in Fig. 2. We have also measured 
the parallatic motion from the Dec to April and then from the April to June, 
both are consistent with each other and with the value given above for
the Dec 2000 to June 2001 total motion.
Also, the individual motion (Dec 2000 to June 2001) in both $\alpha$ and 
$\delta$ are consistent with $4.6 \pm 0.3$~pc within the errors.
This new distance is more precise than previously known
and also consistent with the previous estimates.

\begin{figure}
\vbox{\psfig{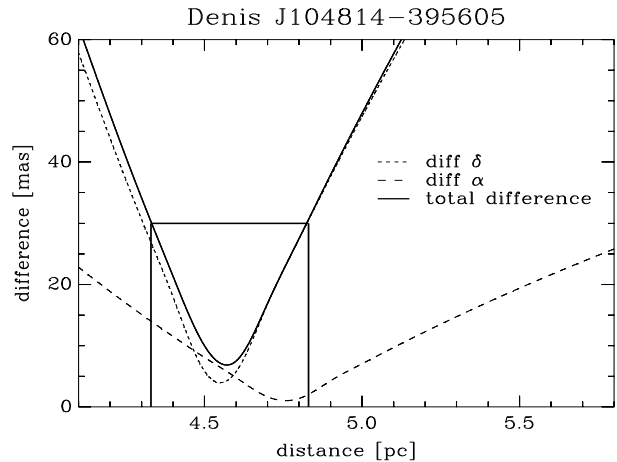}}
\caption{Determining the parallax of Denis~1048$-$39.
We plot the difference (in mas) between observed and expected parallactic motion 
(from Dec 2000 to June 2001) for a range in distance: 
Dashed line for motion in right ascension $\alpha$,
dotted line for declination $\delta$, and full line for total motion.
The difference in total motion is smallest at $\sim 4.6$ pc, 
which is therefore the observed distance towards Denis~1048$-$39.
Given our astrometric precision of 33 mas (see text),
the error is $\pm 0.3$~pc, indicated by the box.}
\end{figure}

\section{Follow-up speckle imaging of Denis~1048$-$39}

In the evening twilight of 2 July 2000, we observed Denis~1048$-$39 using SHARP-I
(System for High Angular Resolution Pictures, Hofmann et al. 1992) at the NTT.
We have obtained a total of 1500 images of 0.5 sec exposure each in the K-band,
spread into six parts of 250 images; from one such part to the next, 
we moved back and forth on two detector quadrants.
According to the La Silla seeing monitor, the average seeing during the
observations was $0.86^{\prime \prime}$.
The FWHM in the final co-added image is $0.34^{\prime \prime}$
(using the nominal SHARP pixel scale of $0.049^{\prime \prime}$).
The data were corrected for bad pixels followed by a sky image subtraction and
the application of a flat-field. For each band we then co-added the $256 \times 256$
pixel frames using the brightest pixel as shift-and-add reference (Christou 1991).

We reduced the SHARP-I speckle data following K\"ohler et al. (2000) to compute
the limiting magnitude for undected but detectable close companion candidates 
from the visibilities, the Knox-Thompson phase, and the bispectrum,
the first of which clearly gives the lowest limits.
The limits for the brightness ratio $K_{1}/K_{2}$ between primary ($K_{1}$) and
any companion candidates ($K_{2}$) for the six data cubes range between 
0.03 and 0.12. When averaging all six cubes, the limit to $K_{1}/K_{2}$ 
is 0.01 for any companion candidate outside of 3 pixels.
We present the achieved dynamic range for the speckle data in Fig. 3,
both for the visibility computed following K\"ohler et al. (2000)
as well as the dynamic range achievable by simple shift and add.
From the visibility curve, we can exclude any companion candidate 
outside of $0.15^{\prime \prime}$ (i.e. 0.7 AU at 4.6 pc) 
down to a limiting magnitude of $\sim 13.6$ mag in K 
(with K$=8.58 \pm 0.05$ mag for the primary, Delfosse et al. 2001),
which corresponds to a $\sim 40$~M$_{\rm jup}$ (Burrows et al. 1997).

\begin{figure}
\vbox{\psfig{figure=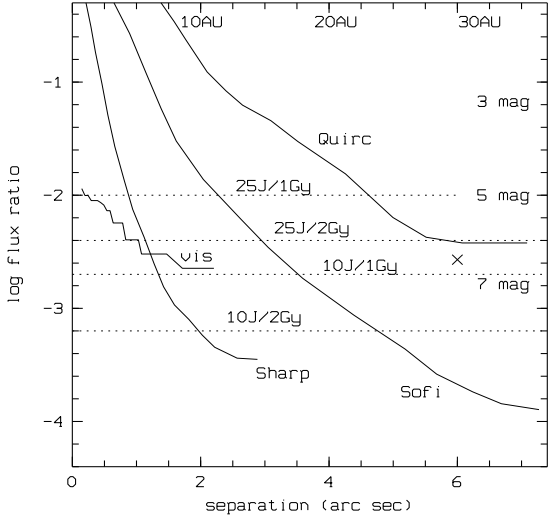,width=12cm,height=8cm,angle=270}}
\caption{Dynamic ranges achieved. We plot the log of the flux ratio between
the $3 \sigma$ background noise level and the peak intensity of Denis~1048$-$39
(and, on the right hand side, the magnitude difference)
versus the separation to the primary's photocenter
(and, on the top axis, the projected physical separation at 4.6 pc),
for the three detectors used (from top to bottom): 
QUIRC at the Mauna Kea 2.2m, Sofi at the La Silla NTT
(both in H) as well as SHARP-I at the NTT (in K);
for SHARP-I, we plot the curve obtained from the simple shift and add image
(from 0 to $3^{\prime \prime}$) and the visivility curve (vis)
calculated following the K\"ohler et al. (2000) proceedure
(for the inner $2^{\prime \prime}$), the latter of which has
better dynanic range for the inner $\sim 1.1^{\prime \prime}$.
The detected companion candidate found to be an unrelated background 
object is indicated as cross ($\sim 2 \sigma$ detection with QUIRC). 
The dotted lines show the expected flux ratios for 10 
and 25~M$_{\rm jup}$ mass companions at 1 and 2 Gyrs around the 
primary (in the H-band) calculated from Burrows et al. (1997) models.
Non-detections of bound companions outside of $\sim 0.25^{\prime \prime}$
(or $2^{\prime \prime}$) exclude any stellar (or any brown dwarf, respectivelly) 
companions outside $\sim 1$ (or 9) AU (at 2 Gyrs). From the SHARP-I
visibility curve, we can exclude companions down to a flux ratio of 1:100
($\sim 40$~M$_{\rm jup}$ at 2 Gyrs) outside of $0.15^{\prime \prime}$ (0.7 AU).}
\end{figure}

The previous companion candidate discussed above is not detected, because we
placed the primary only in the two lower, i.e. northern SHARP-I quadrants, 
because they have better pixel and flat field characteristics;
each quadrant has a field-of-view of $\sim 6.2^{\prime \prime} \times 6.2^{\prime \prime}$.
With the SHARP-I observations, we concentrated on detecting even
closer companion candidates, within $3^{\prime \prime}$; there was 
no reason to image the previously rejected companion candidate again.
No additional companion candidates are detected in the 
$12.5^{\prime \prime} \times 12.5^{\prime \prime}$ SHARP-I field-of-view.
When comparing the PSF of Denis~1048$-$39 with the PSF reference star PPM 288121
(a known single star observed immediately before and after the prime target),
we find no evidence for any elongation of the Denis~1048$-$39 PSF.

In Fig. 3, we compare the dynamic range achieved in our SofI, QUIRC, and SHARP images.
The flux ratio is determined from our actual images of Denis~1048$-$39 as 
the $3 \sigma$ background noise level (on 49 pixels) devided by the peak intensity.
We compare the observed dynamic ranges with expected flux ratios for possible 
companions of different masses (calculated following Burrows et al. 1997),
for an age of 1 and 2 Gyrs for the primary at 4.6 pc distance
(with B.C.$_{\rm H} = 2.6$ mag as for Gl 229 B, Leggett et al. 1999).
Hence, for an age of 2 Gyrs, we should have detected all companions with
mass above $\sim 10$~M$_{\rm jup}$ outside of 5$^{\prime \prime}$ (23 AU)
and all companions with mass $\ge 25$~M$_{\rm jup}$ at $\ge 3^{\prime \prime}$ 
separation (14 AU), all detectable in the SofI images.
The MPE speckle camera SHARP-I clearly gives the best dynamic range.
In the SHARP images, we should have detected all $\sim 10$~M$_{\rm jup}$
companions between 2 and 3$^{\prime \prime}$ separation (9 to 14 AU)
and companions with masses $\ge 25$~M$_{\rm jup}$ at $\ge 1^{\prime \prime}$ 
separation (5 AU).
If Denis~1048$-$39 is a $\sim 0.09$~M$_{\odot}$ mass star (see below), any companion
which is $\ge 1$ mag fainter would be sub-stellar (Chabrier et al. 2000).
Outside of $0.15^{\prime \prime}$ (0.7 AU), we should have detected
all companions with magnitude difference of $\Delta K \ge 5$ mag with SHARP-I.

\section{Discussion and Summary}

A faint, possibly sub-stellar companion candidate 6$^{\prime \prime}$ 
NNW of Denis~1048$-$39 turned out to be a background object.
Because several other groups are actively searching for such sub-stellar
companions around nearby stars, we think that it is important also to
report such a negative result.
Because the primary object itself is almost sub-stellar, 
any real companion which is just a few mag fainter, would be sub-stellar.
Because of the high dynamic range achieved with the speckle camera SHARP-I,
we can exclude stellar companions at separations above $0.25^{\prime \prime}$,
which is only 1 AU at 4.6 pc distance, and we can also exclude brown dwarf
companions at $\ge 2^{\prime \prime}$ (9 AU), all for an age of 2 Gyrs.

Three epoch imaging led to the determination of the parallax of Denis~1048$-$39
corresponding to a distance of $4.6 \pm 0.3$ pc.
This distance estimate is consistent with both the photometric distance of $4.1 \pm 0.6$ pc
given in Delfosse et al. (2001) in case that the object is a single main sequence
M9-dwarf as well as with the trigonometric parallax measurement (from archived
plates) by Deacon \& Hambly (2001). 
At this distance, its apparent magnitudes 
(assuming no interstellar reddening)
%
%
the following absolute magnitudes, rarely available for an M9-dwarf:
M$_{\rm B}=20.69 \pm 0.33$, M$_{\rm R}=17.39 \pm 0.24$, M$_{\rm I}=14.36 \pm 0.14$, 
M$_{\rm J}=11.28 \pm 0.14$, M$_{\rm H}=11.24 \pm 0.34$, and M$_{\rm K}=10.27 \pm 0.14$ mag.

At the given distance, its colors and spectral type place the object into the
H-R diagram, so that we can compare its position with theoretical models
(Burrows et al. 1997, Chabrier et al. 2000). The object
is a very low-mass star with a mass of $\sim 0.075$ to $\sim 0.09$~M$_{\odot}$
and an age of $\sim 1$ to 2 Gyrs.
There is no star known that is both more nearby and cooler than Denis~1048$-$39.
At a distance of $\sim 4.6$~pc, it is the 
40th nearest star.\footnote{see http://www.chara.gsu.edu/RECONS/TOP100.htm}

Speckle imaging with SHARP has revealed no additional closer companion candidates,
so that it seems likely that Denis~1048$-$39 is a single star. 
However, an even lower mass or even closer companion cannot be excluded.

Our direct imaging observations of the companion 
candidate at $\sim 6^{\prime \prime}$ separation with an H-band magnitude 
difference of $\sim 6.4$ mag shows that such faint possibly sub-stellar companions 
can be detected directly around nearby ultra-cool dwarfs,
in this case at a projected physical separation of only $\sim 27$~AU, if it were bound.
In the near future, this target and other similar nearby ultra-cool dwarfs
can be observed with the repaired NICMOS at the HST and CONICA-NAOS at the VLT, 
which provides an IR wavefront sensor. Then, we can investigate the multiplicity 
of such objects with even higher dynamic range and higher sensitivity.

\acknowledgements
We would like to thank the ESO User Support Group for assistance, 
the ISAAC team for the VLT service mode observations, and the NTT team with
O. Hainaut, L. Vanci, G. Martin, and J. Miranda for support during
the SofI observations. We are gratefull to Klaus Bickert and 
Rainer Sch\"odel for their help with the Sharp run.
We also thank Eduardo Mart\'\i n for 
introducing Denis~1048$-$39 to us shortly after its publication
on astro-ph and in a press release in November 2000.
RN and WB would like to thank the Institute for Astronomy of
the University of Hawai'i for hospitality, where they have
done part of this research, as well as for allocation of
several observing nights with its 2.2m telescope on Mauna Kea.
We also thank Susan Parker and Andrew Pickles for their help 
with those observations.
RN wishes to acknowledge financial support from the
Bundesministerium f\"ur Bildung und Forschung through the
Deutsche Zentrum f\"ur Luft- und Raumfahrt e.V. (DLR)
under grant number 50 OR 0003.
The authors wish to extend special thanks to those Hawai'ians,
on whose sacred mountain we worked (without having asked). 

{}

\end{document}